# Nonlocal Quantum Information Transfer Without Superluminal Signalling and Communication

Jan Walleczek[1] · Gerhard Grössing[2]

**Abstract** It is a frequent assumption that—via superluminal information transfers—superluminal signals capable of enabling communication are necessarily exchanged in any quantum theory that posits hidden superluminal influences. However, does the presence of hidden superluminal influences automatically imply superluminal signalling and communication? The non-signalling theorem mediates the apparent conflict between quantum mechanics and the theory of special relativity. However, as a 'no-go' theorem there exist two opposing interpretations of the non-signalling constraint: foundational and operational. Concerning Bell's theorem, we argue that Bell employed both interpretations, and that he finally adopted the operational position which is associated often with ontological quantum theory, e.g., de Broglie–Bohm theory. This position we refer to as "effective non-signalling". By contrast, associated with orthodox quantum mechanics is the foundational position referred to here as "axiomatic non-signalling". In search of a decisive communication-theoretic criterion for differentiating between "axiomatic" and "effective" non-signalling, we employ the operational framework offered by Shannon's mathematical theory of communication, whereby we distinguish between Shannon signals and non-Shannon signals. We find that an effective non-signalling theorem represents two sub-theorems: (1) Non-transfer-control (NTC) theorem, and (2) Non-signification-control (NSC) theorem. Employing NTC and NSC theorems, we report that effective, instead of axiomatic, non-signalling is entirely sufficient for prohibiting nonlocal communication. Effective non-signalling prevents the instantaneous, i.e., superluminal, transfer of message-

✉ Jan Walleczek
walleczek@phenoscience.com

Gerhard Grössing
ains@chello.at

[1] Phenoscience Laboratories, Novalisstrasse 11, 10115 Berlin, Germany

[2] Austrian Institute for Nonlinear Studies, Akademiehof, Friedrichstr. 10, 1010 Vienna, Austria

encoded information through the controlled use—by a sender-receiver pair —of informationally-correlated detection events, e.g., in EPR-type experiments. An effective non-signalling theorem allows for nonlocal quantum information transfer yet—at the same time—effectively denies superluminal signalling and communication.

## 1 Introduction

Bell's theorem proved that no quantum theory based on the joint assumptions of "causality and locality" can successfully reproduce the predictions that are yielded by orthodox quantum mechanics [1]. Consequently, for the case of EPR-type nonlocal correlations at space-like distances, Bell's proof implies that the correlations are either (1) beyond physical explanation, at least in locally causal terms, or alternatively (2) the correlations might be explained by physical processes that are, for example, governed by nonlocal, i.e., superluminal, causal influences. John Bell himself was dissatisfied with the first option of no explanation and famously remarked "that correlations cry out for explanation" [2]. Bell's own pursuit of possible causal explanations in quantum mechanics was reviewed by Norsen [3] whose analysis confirmed that "Bell uses the term 'causality'... to highlight that a violation of this [local causality] condition (by some theory) means that the theory posits non-local causal influences, *as opposed to* mere 'non-local correlations'." However, if one *were* to take seriously the explanatory option that "causal influences do go faster than light", then— Bell [2] noted—one should find disturbing "the impossibility of 'messages' faster than light, which follows from ordinary relativistic quantum mechanics...". He concluded that, therefore, for anyone proposing an approach to quantum mechanics based on nonlocal, i.e., superluminal, influences, the "exact elucidation of concepts like 'message' and 'we' ... would be a formidable challenge" [2]. As we will discuss at length further below, Bell here refers to the special, and still incompletely understood, 'role of us'—human observers and of epistemic agents in general—both in the performance of quantum-based experiments and in possible definitions of the non-signalling constraint (e.g., [4]).

Despite these challenges, Bell shared an ongoing interest in de Broglie–Bohm theory even long after publishing his seminal proof (e.g., [5–8]). David Bohm's non-standard formulation of quantum mechanics is well known for positing nonlocally-causal influences as a fundamental, ontic feature of physical reality [9,10]. "This picture, and indeed, I think, any sharp formulation of quantum mechanics", Bell explained in reference to Bohm's theory, "has a very surprising feature: the consequences of events at one place propagate to other places faster than light. This happens in a way that we cannot use for signalling" [7]. The impossibility to signal by way of superluminal influences Bell took to be the central issue for any explanatory, causal approaches to quantum mechanics [7]: "For me then this is the real problem with quantum theory: the apparently essential conflict between any sharp formulation and fundamental relativity." The fact that—50 years after Bell's theorem— this conflict remains unresolved would likely not have surprised Bell who had predicted decades ago [7]: "It may be that a real synthesis of quantum and relativity theories requires not just technical developments but radical conceptual renewal."

The present work offers a communication-theoretic analysis of the conceptual impasse that exists between (1) the *possibility* of superluminal influences and (2) the *impossibility* of superluminal signalling as required by special relativity: Does the presence of superluminal influences necessarily imply superluminal signalling and communication? We will present an answer based on an informational approach in reference to Shannon's mathematical theory of communication [11,12]. Specifically, the present work introduces a conceptual framework for defining, in a technically consistent manner, the difference between signalling, information transfer, and message communication. These concepts are sometimes used interchangeably and often without clear definition in the literature on quantum foundations. We suggest that in discussions of the non-signalling constraint, and of the relationship between signal transmission, causal influences, and information transfers, this lack of definition is largely responsible for the articulation of conflicting positions.

### 1.1 Does Nonlocal Information Transfer Automatically Imply Superluminal Signalling?

There are many instances in the scientific literature where the concept of information transfer is identified directly with the concept of signalling and communication without justification based upon sound principles. Concepts like 'hidden signalling' or 'hidden communication' have been employed, for example, in negative assessments of ontological quantum theories, without communication-theoretic definition of these concepts (e.g., [13–15]). For example, Gallego et al. [15] claimed that "… Bohm's theory is both deterministic and able to produce all quantum predictions, but it is incompatible with no-signalling at the level of hidden variables." Obviously, that claim contradicts the view held by those who are in support of the theoretical possibility of de Broglie–Bohm theory, including the view held by John Bell (e.g., [5–8]). A new conceptual foundation may be required to be able to move beyond entrenched positions. We suggest that the conflicting positions can be traced to the singular fact that contradictory interpretations of the non-signalling theorem have been used: whereas "axiomatic non-signalling" completely disregards the role of scientific observers, i.e. of epistemic agents, an "effective non-signalling" constraint, as described, analyzed, and defined here, takes into account the essential role of epistemic agents in communication and signalling processes.

Better insight into the difference between concepts like 'nonlocal influence' and 'nonlocal signalling' is needed also in light of the following development: Harrigan and Spekkens [16] defined the distinction between $\psi$-ontic and $\psi$-epistemic approaches to quantum theory. That distinction sparked a new wave of work drawing attention again (i) to the question concerning the *reality* of the quantum state (e.g., [17–20]), and (ii) to possible contributions ontological theories might make to a future understanding of quantum theory. For an extensive review and analysis of current developments at this research frontier see Leifer [21]. Not surprisingly, a key question that has resurfaced in this context is Bell's original question as to why superluminal influences cannot be used to transmit superluminal signals. For example, Wood and Spekkens [22] reconsidered the possibility of "quantum causal" explanations, including the problem of why $\psi$-ontic quantum theories such as de Broglie–Bohm theory appear to depend on the strict "fine-tuning" of causal parameters unless such theories are permitted to violate the non-signalling theorem.

What then is the valid interpretation of the non-signalling theorem? Looking ahead, the conceptual framework offered by Shannon's theory of communication processes allows us to distinguish between two types of signals. That distinction may provide workers in quantum foundations with a fresh approach towards analysing the difference between an axiomatic as opposed to an effective non-signalling theorem. The respective signals we will identify as (communication-theoretic) 'Shannon signals' and (signal-theoretic) 'non-Shannon signals'. Summarizing, an *axiomatic* non-signalling theorem denies transmission of Shannon *and* non-Shannon signals alike. Instead, a theorem of *effective* non-signalling denies only transmission of Shannon signals but does *not* prohibit transmission of non-Shannon signals. We will next present an overview of the contrasting uses of the non-signalling theorem in quantum mechanics, employing as a historical reference the two interpretations used by John Bell.

## 2 The Two Interpretations by John Bell of the Non-signalling Theorem

The non-signalling theorem is widely agreed to represent a general 'no-go' theorem in quantum mechanics. However, as was mentioned already, there exist two opposing interpretations of the non-signalling theorem: a *foundational* interpretation as opposed to an *operational* one. As a 'no-go' theorem in quantum mechanics, how is the non-signalling constraint to be validly interpreted? Again we can turn to Bell for insight and guidance because, as we will explain, he employed both interpretations—foundational and operational—at different times. First, in Sect. 2.1, we will argue that the tradition of tacitly identifying potential ontic influences or causes with signalling, in the context of the predictions of Bell's theorem, started with Bell himself [1]. This is the position we have introduced above as "axiomatic non-signalling". Second, in Sect. 2.2, we will show that —in years following publication of Bell's theorem—Bell took up an explicitly operational interpretation of the non-signalling theorem, an interpretation that views signals and messages as operationally distinct from either causes or influences.

### 2.1 Foundational, Axiomatic Interpretation of the Non-signalling Theorem

In his celebrated paper of 1964, Bell tacitly adopted a *foundational* (axiomatic) interpretation of non-signalling [1]: "In a theory in which parameters are added to quantum mechanics to determine the results of individual measurements, without changing the statistical predictions, there must be a mechanism whereby the setting of one measuring device can influence the reading of another instrument, however remote. Moreover, the signal involved must propagate instantaneously, so that such a theory could not be Lorentz invariant." Evidently, Bell here takes the fact that "the setting of one measuring device can influence the reading of another instrument…" as evidence for instantaneous, i.e., *nonlocal, signalling*—"the signal involved must propagate instantaneously". The possibility in a theory of nonlocal, i.e., superluminal, signalling and communication is in direct conflict, of course, with special relativity, and indeed Bell notes that "such a theory could not be Lorentz invariant." However, without further explanation, Bell [1] directly equates the superluminal *influence* with a superluminal *signal*. We have shown before that loss of operational distinction between 'influence' and 'signal' is the mark of foundational, ontic, or *axiomatic*, interpretations—to use the present terminology—of the non-signalling theorem [23]. There is wide agreement that the foundational interpretation, i.e., axiomatic non-signalling, is the interpretation associated with orthodox quantum theory. Before continuing with the description of the lesser known, operational interpretation of non-signalling in Sect. 2.2, we will review briefly how the concept of axiomatic non-signalling has recently been employed as an *apparently* conclusive argument against both the possibility of determinism and nonlocal hidden-variables approaches in quantum mechanics.

### 2.1.1 Axiomatic Non-signalling as an Argument Against Determinism in Quantum Mechanics

An axiomatic non-signalling concept was employed in attempts to question the viability of *any* kind of deterministic approaches to quantum mechanics, challenging the possibility of de Broglie–Bohm theory for example (e.g., [15,24,25]). We have previously referred to such attempts as *generalizations* of Bell's theorem [23]. The question concerning determinism has also come up again in the context of $\psi$-ontic and $\psi$-epistemic approaches to quantum theory (e.g., [16,21]): Can the unpredictability of EPR-type nonlocal correlations count as conclusive evidence in favour of the existence in nature of objective chance, i.e., of *intrinsic* randomness? On the one hand, the objective nature of quantum indeterminism has long been taken for granted and thus represents a key metaphysical assumption of orthodox quantum theory. On the other hand, it must be acknowledged that proof of absolute indeterminism is impossible as a matter of principle—whether by experiments or by mathematical analyses. The impossibility-of-proof argument was addressed by us before and this argument will not be restated here (see [23]). In the hope of by-passing these fundamental experimental and mathematical constraints, it was argued by others that—nevertheless—there might still be a way to decide between the two competing assumptions: determinism or indeterminism? Specifically, it was claimed that the assumption of (axiomatic) non-signalling suffices to eliminate the possibility of determinism at the level of the quantum (e.g., [15,24,25]). In response to that claim, we have noted before that the possibility of determinism cannot be denied on account of non-signalling because the claim rests on the independent validity of three *interdependent* assumptions [23]. Figure 1 illustrates the relational interdependency of the three assumptions that underlie the reasoning behind axiomatic non-signalling as a suggested proof of quantum indeterminism, i.e., of intrinsic randomness, in nature.

We suggest that there does not exist at present a conclusive argument derived from an axiomatic non-signalling assumption which is capable of the successful generalization of Bell's theorem (for details see [23]). A related but not identical argument was previously offered by Ghirardi and Romano [19]. To repeat, it remains undecidable,

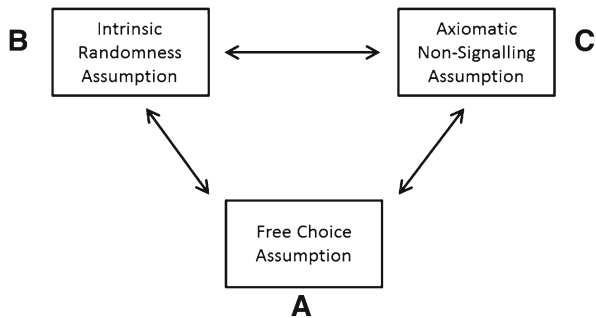

**Fig. 1** Relational diagram illustrating the irreducible interdependency of basic metaphysical assumptions implicit in standard interpretations of orthodox quantum theory (adapted from [23]). **a** Free choice assumption, **b** intrinsic randomness assumption, and **c** axiomatic non-signalling assumption. Crucially, the validity of interpreting the non-signalling theorem as an ontic, foundational theorem, or axiom, for quantum mechanics, a frequent assumption in standard interpretations, depends on the independent validity of assumptions (**a**) and (**b**). However, neither assumption (**a**) nor assumption (**b**) can be *independently* confirmed if the possibility of 'free choice' depends on the existence of a process that is intrinsically random and *vice versa*

on the basis of logical considerations alone, which of the opposing metaphysical positions is valid—indeterminism or determinism (see Fig. 1). In the following, we will explain that Bell distanced himself from the foundational, axiomatic interpretation and that he started to adopt an operational approach in line with a theorem of *effective* non-signalling [26].

### 2.2 Operational, Effective Interpretation of the Non-signalling Theorem

In the years following publication of his ground-breaking theorem, Bell introduced work that recognized and upheld in quantum theory the operational distinction between 'influences' or 'causes', and 'signals' or 'messages'. Specifically, Bell considered the concept of potential ontic influences in quantum mechanics, i.e., the concept which he introduced as 'beables' by way of contrast to the standard concept of "observables", which refers to (epistemic) states of knowledge only [26]. Significantly, these (potentially ontic) 'beables', he classified into "controllables", and "uncontrollables", whereby he noted that the "latter are no use for *sending* signals..." [26]. Note that Bell's distinction between "controllables", and "uncontrollables" does not inherently depend on the distinction between nonlocal and local beables, or whether beables are unpredictably deterministic or intrinsically random. In summary, starting in 1976, Bell's own work introduces an *operational* interpretation of the non-signalling constraint, i.e., to use our present terminology, Bell shifts his attention from an axiomatic, towards an *effective*, non-signalling concept (e.g., [5–8, 26, 27]).

Lately, there has been a resurgence of interest in Bell's own interpretation of *his* theorem before and after the year 1976. Wiseman [28, 29] recently considered the idea of "The two Bell's theorems of John Bell". The shift in Bell's own thinking could be accounted for by a novel perspective on the non-signalling constraint, a view that is implied by Wiseman's analysis also [29, 30]. We here argue that the transition from Bell's position of 1964 [1] to that of 1976 [26] represents a movement towards an operational approach which specifically seeks to account for the effective role of *epistemic agency*, i.e., agent *control* based upon agent *knowledge*, as a key factor in the construction of the non-signalling constraint (see Sect. 4.2). Again, as was mentioned already, the agent-based view of non-signalling is well-exemplified by Bell's own assertion, that "we cannot use for signalling" *the way* in which "events at one place propagate to other places faster than light" [7]. Obviously, Bell' s new view of (effective) non-signalling is in stark contrast to his original position of (axiomatic) non-signalling [1].

Historically, the number has been growing of researchers who have argued—in one form or another, directly of indirectly—for the validity of an agent-based view of the non-signalling theorem in line with Bell's notion of operationally "uncontrollables" of 1976 (e.g., [3,19,20,23,28–35]). Note that this list is far from complete. What has been lacking so far, however, is an understanding of the communication-theoretic difference between axiomatic and effective approaches beyond the statement that agent participation is presumed in the effective approach. The present work seeks to identify a decisive technical criterion for defining the difference between "axiomatic non-signalling" and "effective non-signalling" in the context of communication theory.

### 2.2.1 An Effective Non-signalling Theorem Accounts for the Controlling Actions of Epistemic Agents

That Bell saw as indispensable the special role of the experimenter agent in reaching a full understanding of the non-signalling theorem is amply evident in Bell's last published statement on this matter [8]: "Do we then have to fall back on 'no signalling faster than light' as the expression of the fundamental causal structure of contemporary theoretical physics? This is hard for me to accept. For one thing we have lost the idea that correlations can be explained, or at least this idea awaits reformulation. More importantly, the 'no-signaling...' notion rests on concepts which are desperately vague, or vaguely applicable. The assertion that 'we cannot signal faster than light' immediately provokes the question: Who do *we* think we are? *We* who can make measurements, *we* who can manipulate 'external fields' , *we* who can signal at all, even if not faster than light? Do we include chemists, or only physicists, plants, or only animals, pocket calculators, or only mainframe computers?". Note, that—by contrast—the consideration of who or what qualifies as an epistemic agent, or whether agents may, or may not, access controllably information transfers—plays no role at all in a theorem describing axiomatic non-signalling. We will return to the concept of knowledge-based agents in quantum models expressing effective non-signalling in Sect. 4.2. Next, we will introduce our informational approach for answering Bell's quantum-foundational questions regarding the non-signalling notion.

## 3 An Informational Approach Towards the Concept of Hidden Superluminal Influences

One major reason as to why many researchers insist on the foundational, or axiomatic, interpretation of the non-signalling theorem often comes from the following under-standing: The presence of hidden superluminal influences would necessarily imply the presence of (hidden) information transfers which in turn would imply signal transfer and the possibility of communication. It is a frequent assumption that—via superluminal information transfers—superluminal signals *capable of enabling communication* are necessarily exchanged in any quantum theory that posits superluminal influences (compare Sect. 1.1). As a consequence, relativity theory would be violated which would render an ontological quantum theory, like de Broglie–Bohm theory, physically unrealistic. The analysis provided in Sect. 4 will explore whether or not this understanding is justified by definitions of signalling, information transfer, and message communication, based upon Shannon's theory of communication processes [11,12].

The ensuing analysis will not only be consistent with the informational approach towards analysing various consequences of proposed superluminal (causal) influences but our analysis will strictly rely on that approach to draw its final conclusions. That is, we assume from the start that "hidden superluminal influences" (e.g., [14]), and the possibility of nonlocally-causal transfers (e.g., [9,10]), invariably involve information transfers or exchanges. For example, Bohm and Hiley [31] used the term "active information" to describe such nonlocal exchanges in the context of nonlocal hidden-variables approaches. We here take a 1-bit informational event to be the minimal indication for the occurrence of any kind of causal exchange, or any discernible physical influence. More precisely, without detection of nonlocally-correlated informational events—between two space-like separated members of an entangled pair—evidence for any (potentially ontic) influence between pair members would be entirely unavailable. Evidently, then, the central question here considered reduces to this: Must information transfer always imply signalling and communication? Specifically, how could signal exchanges between two systems be denied if—at the same time—unrestricted informational exchanges are allowed between them? Following the detailed communication-theoretic analysis of these questions in Sect. 4, we will consider the application of the subsequent definitions in the context of an effective interpretation of the non-signalling theorem for quantum mechanics in Sect. 5.

## 4 Information Transfer, Signal Transfer, and Message Communication, in Reference to Shannon's Mathematical Theory of Communication

To begin with, we informally define two theorems that are implicit in Shannon's theoretical framework, i.e., theorems that remain unstated usually because they appear to be self-evident. For the present work, however, we make explicit these theorems by naming them: (1) a theorem of information transfer control (ITC), and (2) a theorem of information signification control (ISC). The effective role and application of these theorems in the context of constructing an effective non-signalling theorem in relation to quantum mechanics will be described in Sect. 5. Importantly, these normally *implicit* theorems provide the larger *operational* context without which Shannon's familiar *explicit* theorems and mathematical measures such as Shannon's source coding theorem or Shannon's channel information capacity, C, could not be usefully applied in any practical manner [11]. In other words, the normally implicit theorems account for the involvement of epistemic agents who originate, encode, send, receive, and decode, signals. When assessing—in the context of communication theory—the relationship between the non-signalling theorem, quantum theory, and special relativity, we propose that Shannon's implicit theorems provide a consistent operational framework, because ITC and ISC theorems account specifically for agent participation during signalling and communication processes. Subsequent Sect. 4.1 will provide an explanation of essential differences between (i) Shannon's *operational* framework which includes, for example, agent-dependent encoding/decoding processes as accounted for by the ISC theorem, and (ii) Shannon's *mathematical* framework which is independent of agent participation.

### 4.1 Shannon's Information-Theoretic Approach to Human and Automated Machine Communication

We next introduce Shannon's well-known concepts [11]: "The fundamental problem of communication is that of reproducing at one point either exactly or approximately a message selected at another point." For Shannon, any possible communication starts with an "*information source* which produces a message or sequence of messages to be communicated to the receiving terminal". More precisely, the message to be communicated originates with a '*discrete*' information source. "We can think of a discrete source", Shannon [11] explained, "as generating the message, symbol by symbol. It [the source] will choose successive symbols according to certain probabilities depending, in general, on preceding choices as well as the particular symbols in question."

To illustrate the concept of Shannon's 'discrete source' , Fig. 2 shows the scenario of bidirectional communication of the message "SOS"—in Morse code—between two sources here represented by sender Alice and receiver Bob. It is apparent that Shannon's discrete source for "generating the message", e.g., Alice in Fig. 2, manifests *operational control* over the process of signal transmission through the *discrete channel*. Shannon defined a 'discrete channel' as a "... system whereby a sequence of choices from a finite set of elementary symbols $S_1, \ldots, S_n$, can be transmitted from one point to another." The sequence "• • • − − − • • •" shown in Fig. 2 was chosen "symbol by symbol" by Shannon's information source, e.g., epistemic agent Alice, from the "set of elementary symbols" offered by Morse code (•, −) to communicate the message "SOS". The point is that a *random* sequence of symbols could *not*—of course—effectively transmit the message "SOS" between Alice and Bob.

In contrast to the scenario in Fig. 2, which illustrates Shannon's *operational* framework, Shannon's *mathematical* framework for describing channel information capacity, C, is founded upon an entirely different assumption about the properties of an information source. "How is an information source to be described mathematically", Shannon [11] asked, and he introduced the concept of a *random* source to account quantitatively for the transmission of any *arbitrary* message from a set of *possible* messages. As was emphasized by Shannon [11], the meaning of an *actual* message, and thus of any "successive symbols", is of no importance in calculating C: "Frequently the messages have *meaning*; that is they refer to or are correlated according to some system with certain physical or conceptual entities. These semantic aspects

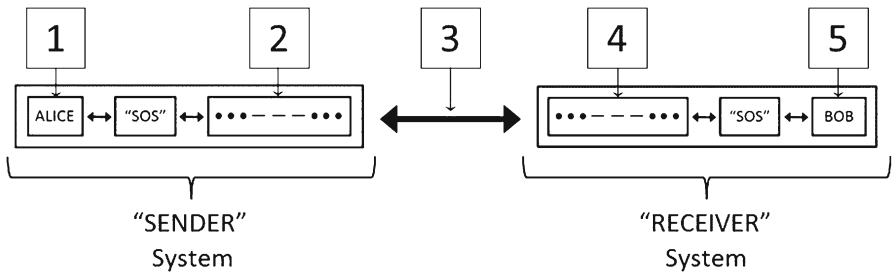

**Fig. 2** Illustration of basic concepts from Shannon's information-theoretic approach to sender–receiver systems in human and automated machine communication [11]. For easy overview, we describe key functional elements that operationally define Shannon's view of sender-receiver systems, employing his original definitions [11]: (1) An epistemic agent, here named 'Alice', is the "source which produces a message", e.g. the three-letter message "SOS"; (2) an encoding transmitter "which operates on the message in some way to produce a signal suitable for transmission" by performing—here in (binary) Morse code—"an encoding operation which produces a sequence of dots, dashes, and spaces ... corresponding to the message"; (3) a conduit or channel, i.e., "the medium used to transmit the signal from transmitter to receiver"; (4) a decoding receiver which "performs the inverse operation of that done by the transmitter, reconstructing the message from the signal"; finally (5), an epistemic agent, here named 'Bob', is the destination "for whom the message is intended". We label combination of elements (1) and (2), i.e., the epistemic agent and the transmitter, a "sender" system, or simply the "sender". Accordingly, we label combination of elements (4) and (5), i.e., the receiver and the epistemic agent, "receiver system", or simply the "receiver". The bi-directionality of arrows in the figure serves to indicate that the two systems, "sender" and "receiver", each may perform the functions of the other leading to the possibility of bidirectional message communication

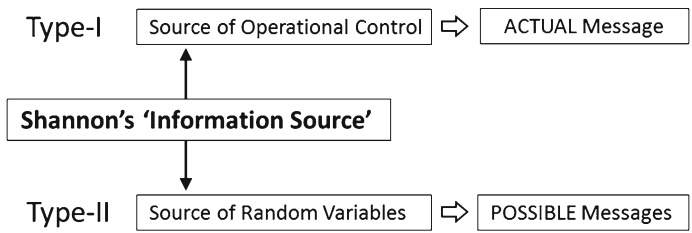

**Fig. 3** The two types of 'information source' in Shannon's communication theory. The type-I source represents a source of operational control capable of generating an *actual* message for the purpose of sender-receiver communication (e.g., see the message "SOS" in Fig. 2). The type-II source represents a source of random variables, i.e., a "stochastic process" [11], that probabilistically generates the set of *possible* messages in the mathematical quantification of channel capacity C

of communication are irrelevant to the engineering problem. The significant aspect is that the actual message is selected from a set of possible messages." In reference to Shannon's distinction between *actual message* and *possible messages*, Fig. 3 represents two different types of 'information source' which we have named type-I and type-II; both types are respectively employed in Shannon's model of communication [11].

The so-called "engineering problem" refers to the *mathematical account* of channel information-carrying capacity for theoretically *possible* messages. By contrast, the problem which is the focus of the present work concerns the *operational account* of how an *actual* message is transmitted from sender to receiver such as the message

"SOS" in Fig. 2. Importantly, each type of problem—actual message *versus* possible messages—elicits different properties of an 'information source' as illustrated in Fig. 3. In Shannon's *operational* framework of *actual* message communication the participation of epistemic agents who represent *type-I sources* manifesting operational control is strictly required (Fig. 3). That is, type-I agents choose—to use Shannon's words again—"successive symbols... depending, in general, on preceding choices as well as the particular symbols in question" (compare Fig. 2). It is evident that Shannon's use of the term 'choice' does *not* refer to a *random* process here. Instead, in the context of generating an *actual* message, Shannon's 'information source' represents a *source of operational control* (see 'type-I source' in Fig. 3 and Sect. 4.2). The corresponding physical signals, i.e., those passing through the channel under type-I agent control, we from now on will refer to by the new term 'Shannon signals' (see Sect. 4.3 for details). Importantly, the above-mentioned ITC and ISC theorems constrain only the transmission of such *Shannon signals*, i.e., the signals that deliver *actual* messages (see Sects. 4.3.1 and 4.3.2).

By contrast, for the *mathematical* description of the "set of possible messages" Shannon's information source represents a *source of random variables* (see 'type-II source' in Fig. 3). The type-II source represents—by Shannon's description—a "stochastic process" for the purpose of producing probabilistically the set of *possible* messages in the quantification of channel capacity [11]. Importantly, unlike the assumptions A and B described in Fig. 1 (see Sect. 2.1.1), the (type-II) source of random variables in Shannon's theory can be *deterministic*. An example of the physical instantiation of a type-II source would be the tossing of a fair coin revealing either 'heads' or 'tails', which is a process that is deterministic yet operationally *unpredictable*.

This work focusses on the physical processes that facilitate *actual* message communication between epistemic agents, e.g., the controlling actions of agents described by *type-I* operations (see Fig. 3). It is important to note that there *cannot* be—in actuality—'random messages', i.e., random acts of *actual* communication between sender and receiver; there could only be random *information transfers* as will be explained in detail in Sect. 4.3. Who or what is the epistemic agent in relation to Shannon's theory, Bell's theorem, and the non-signalling constraint?

### 4.2 Defining the Epistemic Agent

As was mentioned in Sect. 2.2.1, John Bell was keenly aware of the importance in interpretations of the non-signalling theorem of having an understanding of *agency*. Who or what represents an (epistemic) agent? Remember that Bell once asked whether we should include in our definition "... chemists, or only physicists, plants, or only animals, pocket calculators, or only mainframe computers?" [8]. We will here introduce a definition that is intended primarily to clarify the meaning of his question, and that may direct us towards possible answers in the spirit of Bell's original inquiry. Agency is generally defined as the capacity of humans or other entities *to act in the world*. Put differently, an agent is defined initially by possessing the capacity to influence causal flows in nature. By prefacing "agent" with the term "epistemic", attention is drawn to the fact that a complete definition of agency represents more than the mere "capacity to influence causal flows": an agent possesses knowledge-based, i.e., epistemic, capacity for *predictably* directing, and redirecting, causal flows, and thus for directing, and redirecting, *information* flows as well. That is, an epistemic agent holds the power to (statistically) control physical activity based upon an ability to predict the outcome of specific actions on targeted processes in reference to a known standard or goal. In short, an epistemic agent thus manifests in the world a genuine source of *operational control* (see Fig. 3). By the definition here introduced, Alice and Bob in Fig. 2 represent *type-I* agents who each may plausibly represent and enact in Shannon's sender-receiver system an "information source which produces a message" (see Fig. 3). It is thus apparent that message communication, in the sense of Shannon's communication theory, strictly requires the participation of epistemic type-I agents who can originate, encode, send, receive, and decode, signals (compare Fig. 3). What exactly constitutes "signalling", and "non-signalling", in the context of the present communication-theoretic investigation of the non-signalling theorem?

### 4.3 The Communication-Theoretic Distinction Between Shannon Signals and Non-Shannon Signals

We have before introduced the new concept of 'Shannon signal' in Sect. 4.1. What exactly constitutes a 'signal'? Unless one finds agreement first on what represents a signal, one cannot later expect to have agreement on an appropriate definition of 'non-signalling', i.e., negation of signalling. We next distinguish between two types of signals that are apparent in the context of Shannon's theory: (1) signals in the familiar sense of standard *signal theory*, and (2) signals in the sense of *communication theory* only. The latter type of signals we have referred to as "Shannon signals" whereas the former we will refer to as "*non*-Shannon signals". For explanation, the notion of non-Shannon signal represents a signal in the standard (signal-theoretic) sense of detecting a physical influence as part of measurement processes in general, *independent* of an effective communication task. A simple engineering example is the manifestation of a "click" by a suitable threshold detector in response to physical stimulation. However, this is not what is exclusively meant by the term 'signal', or 'signalling', in the context of Shannon's communication theory. There, the concept of signal also refers to the *controlled delivery* of an informational bit sequence (or, alternatively, a single bit of information) which has been subjected to a process of *message encoding* (by Shannon's 'encoding transmitter'; see Fig. 2). In agreement with that use in Shannon's theory, we informally define 'signal' in the communication-theoretic sense as "controllably-transmitted and message-encoded information". Summarizing, a non-Shannon signal represents a signal in the well-known sense of standard signal theory, whereas the newly introduced concept of Shannon signal finds application only in the communication-theoretic context provided by Shannon's model.

Next, different communication-theoretic scenarios will be compared to explain further the difference between Shannon signals and non-Shannon signals. The scenarios demonstrate negation of Shannon signal transmission while—at the same time— allowing transmission of non-Shannon signals, e.g., of uncontrollably-generated

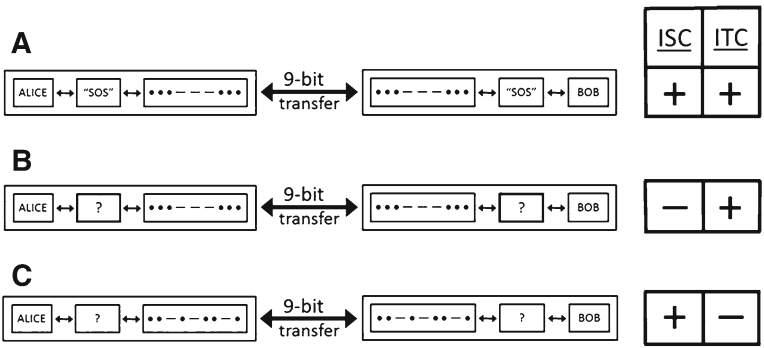

**Fig. 4** Illustration of 9-bit information transfer comparing 'signalling information transfer' (**a**) and 'non-signalling information transfer' (**b**, **c**). The *boxes* to the far right of the figure indicate when is available (+) or not (−) to epistemic agents Alice and Bob the capacity of *ISC* (information signification control) or *ITC* (information transfer control) in a given scenario (**a**–**c**). Question marks shown inside *smaller boxes* (see **b**, **c**) indicate that—even though perfect informational correlations can be observed and recorded at Alice's and Bob's locations—the communication of the message "SOS" is effectively denied. For example, although Bob obtains perfect knowledge about the informational state of Alice's transmitter upon simply observing the informational state of his receiver, he nevertheless is denied reception of any message from Alice (see **b** and **c**). The examples in **b** and **c** demonstrate that the mere fact alone of the availability to Bob and Alice of informational correlations at their respective locations—as part of some communication system—need *not* at all indicate the presence of a message or signal in the communication-theoretic sense. Importantly, this conclusion is entirely independent of the fact whether the involved information transmission channel would be represented by a *quantum* channel or a *classical* channel. In short, while every signal or message represents information, not every information represents a message or signal

informational events. Figure 4 illustrates a total of three scenarios (A–C), including the standard scenario employed in Shannon's theory [11] as a reference (Fig. 4a). Figure 4b and c illustrate the respective applications of the ISC theorem (Sect. 4.3.1) and the ITC theorem (Sect. 4.3.2).

### 4.3.1 The Information-Signification-Control (ISC) Theorem

In Shannon's informational model of signalling processes, any signal sequence is composed of discrete and elementary informational (bit) units [11]. Crucially, single bits or bit sequences have no meaning "in-themselves", that is, they do *not* represent Shannon signals or actual messages. Importantly, the process that grants shared meaning, i.e., shared semantic content, to basic syntactic elements, such as an informational bit sequence, is known in semiotic theory as '*signification*' (e.g., [36]). Generally, semiotic theory studies the relationship between *signs* and (the process of creating) *meaning*. For example, a *code* represents a rule for connecting signs, e.g., dots and dashes in the case of Morse code, to their intended meaning. We have introduced "signification" as a technical term in order to describe accurately the process of assigning meaning to information by way of encoding and decoding mechanisms that could be shared between sender and receiver. The concept of ISC makes explicit for Shannon's model the normally implicit theorem that operationally accounts for encoding and decoding processes (see also *Introduction* to Sect. 4). For example, Fig. 4a and b portray identical informational patterns. However, only for the standard scenario discussed by Shannon, the 9-bit pattern represents a *Shannon* signal (i.e., here the signal which—in Morse code—conveys the message "SOS" ; see Fig. 4a). Conversely, the structurally-identical 9-bit pattern, shown as part of the second scenario in Fig. 4b, represents a *non-Shannon* signal. The difference between the two scenarios is explained by the *presence* (Fig. 4a) or *absence* (Fig. 4b) of operational control by epistemic agents over encoding-decoding processes, as accounted for by the ISC theorem. Put simply, *epistemic* agents Bob and Alice do *not know* Morse code in this example and they cannot therefore exchange *actual* messages using this coding system. The crucial point is the following: whereas Shannon's standard scenario represents information transfer that is *signalling* (Fig. 4a), the scenario illustrated in Fig. 4b represents the case where the transfer of information is *non-signalling* in Shannon's communication-theoretic sense. That is, (two-way) transfers of Shannon signals are denied until Alice and Bob each acquires operational control over the process which we refer to as signification.

### 4.3.2 The Information-Transfer-Control (ITC) Theorem

To review the above, in the scenario in Fig. 4b, unlike in the standard scenario in Fig. 4a, operational control was absent by Bob and Alice over ISC (e.g., code or key sharing). By contrast, Fig. 4c illustrates the opposite scenario: control over information signification is available, whereas control over information transfers through the transmission channel is not, even on a statistical basis only (compare boxes to the right of Fig. 4b and c). To illustrate this scenario when agents Bob and Alice lack the power to control information transfers, the sequence "● ● − ● − ● ● − ●" is shown as one possible, unpredictably transmitted informational pattern. This pattern represents, of course, the appearance of non-Shannon signals only, to apply the present terminology (Fig. 4c); naturally, any other informational bit sequence pattern could have also been selected instead to visualize the statistical uncontrollability of information transmission. Note also that uncontrollable processes may *accidentally* generate the pattern "●●● − − − ●●●" shown in Fig. 4b, and may thus potentially transmit a *false-positive* "signal" between Alice and Bob. In summary, the ITC theorem represents the negation of operational control by epistemic agents over information transmission (see also *Introduction* to Sect. 4).

## 5 Two Sub-theorems Represent an Effective Non-signalling Theorem

Why is the identification misleading of "hidden signalling", "hidden messages", or "hidden communication", with the concept of "hidden information transfer", for example, in the context of de Broglie–Bohm theory? To answer that question, we next employ key findings from the previous section in the task of defining the non-signalling theorem as an effective instead of as an axiomatic theorem. Again, the effective view represents the negation of operational control by epistemic type-I agents (see Fig. 3). In accordance with the above communication-theoretic analysis, the *complete* negation of operational control includes two distinct aspects: (i) negation of ITC (see Fig. 4c), and (ii) negation of ISC (see Fig. 4b). Simply on that basis, we distinguish between two sub-theorems that jointly represent an effective non-signalling theorem, and we introduce them as: (1) Non-transfer-control (NTC) theorem (i.e., negation of ITC), and (2) Non-signification-control (NSC) theorem (i.e., negation of ISC). Importantly, in light of the hypothesis of *quantum channels* for instant information transmission, e.g., as postulated in de Broglie–Bohm theory, the following is obvious: while the NSC theorem must be theoretically accounted for in any complete description of an effective non-signalling theorem (compare Fig. 4b), it is immediately apparent also that sender and receiver could, in principle, share knowledge about any arbitrary coding system with each other; however, only *classical channels* could be used for that purpose. Concerning the NTC theorem, operational uncontrollability of information transmission via hypothetical quantum channels assures that nonlocal information transfers are entirely non-signalling—in the sense of denial of (communication-theoretic) Shannon signalling.

To repeat, NTC and NSC theorems negate *Shannon signalling*, i.e., *actual* message transfers, but they do *not* interfere with the transmission of *non*-Shannon signals (for definitions see Sect. 4.3). To show that negation only of Shannon signalling, but not negation of *non*-Shannon signal transfer, is the relevant concept for constructing an effective non-signalling constraint, the following comparison may advance insight: Evidently, the non-signalling theorem was *not* introduced to prohibit the manifestation by quantum detectors—in EPR-type experiments—of (nonlocally-correlated) measurement clicks that reveal informational correlations at space-like distances (e.g., [37–42]). Instead, we maintain that the non-signalling theorem was introduced to prevent the possibility of the instant, i.e., superluminal, transfer of message-encoded information through the controlled use—by a sender-receiver pair—of informationally-correlated detector clicks. While it is certainly true that each individual click represents a signal also in the ordinary signal-theoretic sense (i.e., in the sense of a *non*-Shannon signal as defined in Sect. 4.3), the detection of a click need not at all indicate the presence of a signal as defined in terms of "controllably-transmitted and message-encoded information", i.e., in the sense of the communication-theoretic definition of (Shannon) signal. Consequently, it would be wrong—both in the context of Shannon's theory and the non-signalling theorem—to use as synonyms 'information' and 'signal', and thus to simply identify information transmission with signal transmission and communication (see also legend to Fig. 4). In summary, of the two sub-theorems of an effective non-signalling theorem—NTC theorem and NSC theorem—each one represents a theorem about the negation of 'operational control' (see Sects. 4.3.1 and 4.3.2). Operational uncontrollability of potentially ontic influences, i.e., 'beables', Bell [26] took as the primary justification for adopting an *effective* non-signalling theorem, instead of an *axiomatic* one [1]. Figure 5 provides a summary (i) of the two contrasting interpretations by John Bell of the non-signalling theorem and (ii) of our communication-theoretic criterion which accounts for the difference between Bell's original position of 1964 [1] and his later position which he first introduced in 1976 [26].

This work may represent an important step towards answering Bell's pressing questions which were motivated by his sceptical assessment that "... the 'no-signaling...' notion rests on concepts which are desperately vague, or vaguely applicable" [8]. The present analysis sought to clarify the concepts in question, including of the 'role of us'

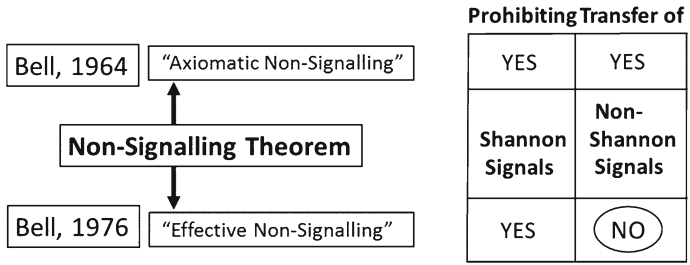

**Fig. 5** The two interpretations by John Bell of the non-signalling theorem (for details see Sect. 2). To the left of the figure are represented the two contrasting interpretations of the non-signalling theorem—the foundational, axiomatic position of 1964 [1] and the operational, effective position of 1976 [26]. To the right of the figure is summarized the communication-theoretic criterion which technically defines the difference between "axiomatic non-signalling" and "effective non-signalling", in accord with the *present* analysis. The boxes at the top ("YES" and "YES" ) indicate that "axiomatic non-signalling" eliminates the distinction between Shannon and non-Shannon signal transfer and thus ignores any generative role for epistemic agents in the establishment of signalling events in the communication-theoretic sense. The boxes at the bottom ("YES" and "NO" ) indicate that "effective non-signalling", by contrast, maintains the communication-theoretic difference between the respective signals: an effective non-signalling theorem prohibits only the (type-I agent-controlled) transfer of Shannon signals but does not prohibit the transfer of non-Shannon signals which is independent of type-I agent control (compare Fig. 3)

—scientific observers and epistemic agents in general—as part of the non-signalling notion, with the following result: *We* are the agents who can manifest control over (i) information transfers as well as (ii) information signification but only to the extent that we are granted access to informational pathways and to shared symbolic representation (see also legends to Figs. 4, 5). Figure 5 illustrates the fact that the foundational, axiomatic interpretation eliminates the distinction between Shannon and non-Shannon signals. Consequently, the axiomatic position on non-signalling fails to address Bell's urgent questions such as about scientific observers "who can signal at all, even if not faster than light?" [8]. An effective interpretation, by contrast, acknowledges the pivotal difference between the respective signals, leading to the possibility of developing a communication-theoretic account of agent participation during signalling and non-signalling processes: an effective non-signalling theorem only limits (type-I agent-controlled) transmission of Shannon signals but does not prohibit transmission of non-Shannon signals (see Fig. 5).

Finally, having in hand now a single criterion for the consistent differentiation between axiomatic non-signalling and effective non-signalling, questions to be explored in future work include the following: Does the proposition of nonlocal information transfer automatically entail a space-time (metric) structure that ceases to be Lorentz invariant? Does the possibility of superluminal information transfer compromise special relativity, even though the possibility of superluminal signalling and communication, by way of instantaneous, i.e., nonlocal transfers, is fully denied? In other words, is it inevitable that causal paradoxes are generated *automatically*—as a function of nonlocal quantum information transfers—even in the complete *absence* of experimenter agents and of their laboratory devices, i.e., *if nature is left to herself*? The subsequent discussion is intended to provide an initial orientation for such work.

## 6 Discussion

The present analysis has identified minimal conditions for an operational non-signalling theorem, i.e., a theorem that allows for the possibility of nonlocal quantum information transfer, yet one that effectively denies superluminal signalling and communication. Our findings may facilitate the translation of ideas and concepts – in the context of quantum-entangled information—between uncontrollability of nonlocal influences and the observation of nonlocal informational events in EPR-type experiments. Furthermore, this work may prove useful for assessing Bell's original notion of "uncontrollables" within the scope of more recent physical, metaphysical, and epistemological interpretations of quantum information and entanglement (e.g., [43,44]). For this discussion the focus will be on three questions, each of which highlights a specific research challenge when interpreting the wave function $\psi$ as a reality, including in nonlocal hidden-variables theories.

### 6.1 What is the Physical Meaning of the Term "Hidden" in the Context of Hidden-Variables Theories?

What is "hidden" about 'hidden variables'? "The usual nomenclature, *hidden variables*, is most unfortunate", Bell [45] complained in a note, and he proposed that "Perhaps uncontrolled variable would have been better, for these variables, by hypothesis, for the time being, cannot be manipulated at will by us." If one accepts Bell's proposal that hidden variables represent *uncontrollable* variables, i.e., variables that "cannot be manipulated at will by us", then much of the mystery is lifted surrounding the term "hidden" : the variables are called "hidden" because they are uncontrollable for any pragmatic purposes, such as for sender-receiver communication. However, the determinism in Bohm's theory, should it not imply statistical predictability and control? The false habit of identifying determinism with predictability stems from an idealized view of physical systems as fundamentally *linearly*-behaving systems: If one knows the initial state of a system then one can predict its evolution towards the final state. However, if there is involved only the weakest element of *nonlinearity* in a deterministic system, such as is the case with sensitive dependencies on initial state conditions, then final state prediction may quickly become impossible (e.g., see deterministic chaos in emergent dynamics). For explanation, even the best nonlinear control and prediction techniques, whether applied to emergent, self-organizing states in physical, chemical, or living systems, are successful only in very limited nonlinear regimes due to the prohibitive complexity of multi-factorial, randomizing interactions (e.g., [46]). "Consider the extreme case of a 'random' generator", Bell [27] explained, ". . . which is in fact perfectly deterministic in nature—and, for simplicity, perfectly isolated. In such a device the complete final state perfectly determines the complete initial state—nothing is forgotten. And yet for many purposes, such a device is precisely a 'forgetting machine'. A particular output is the result of combining so many factors, of such a lengthy and complicated dynamical chain, that it is quite extraordinarily sensitive to minute variations of any one of many initial conditions. It is the familiar paradox of classical statistical mechanics that such exquisite sensitivity to initial conditions is practically equivalent to complete forgetfulness of them." In de Broglie–Bohm theory, just as Bell [27] had suggested with his thought experiment of the "forgetting machine", the *operational impossibility* to predict and control *individual* quantum correlations derives from the impossibility to know—with arbitrary precision—initial state configurations of the *nonlinearly*-behaving, yet fully deterministic, (sub)quantum system (e.g., [31,32]). In fact, from the vantage point of an effective non-signalling constraint, the 'hidden variables' associated with the notion of superluminal hidden influences represent 'non-signalling variables' in the here introduced communication-theoretic sense of (Shannon) non-signalling (see Fig. 4).

### 6.2 Does Nonlocal Information Transfer Nevertheless Violate Relativity Theory Despite the Fact that the Information is Non-signalling in the Communication-Theoretic Sense?

"Does superluminal information transmission automatically violate relativity theory?" asked Maudlin [35]. An affirmative answer is often assumed tacitly by researchers who are committed to the standard, axiomatic view of the non-signalling condition. However, Maudlin's extensive analysis reveals a more complex picture. He concludes that—by itself—the possibility of superluminal information transmission "need not give rise to causal paradox, if the information is not available for general use." Specifically, his analysis finds that the generation of causal paradoxes is the consequence of the possibility of "*signal* loops and where causal processes cannot be used to send signals paradoxes cannot arise." By our definition, Maudlin [35] here refers to the formation of Shannon-signal loops, and if their formation is denied, then causal paradoxes cannot occur. An effective non-signalling theorem, as offered in the present study, denies Shannon signalling and therefore negates Shannon-signal loops from being formed (see Figs. 4, 5). Consequently, the standard assumption may be in need of revision that a quantum theory must be unphysical, e.g., in violation of special relativity, simply because it allows for the possibility of nonlocal information transfer. We here suggest that the effective, operational position on non-signalling may negate the presumed inevitability of paradoxical consequences in association with quantum theories positing nonlocal information transfer. However, in regards to the notion of nonlocal, i.e., superluminal, influences is it acceptable to invoke terms such as 'transfer' or 'transmission' ?

### 6.3 What Justifies the Use of Terms Like 'Transfer' or 'Transmission' in Relation to the Concept of Nonlocal Quantum Information?

In what sense is it acceptable to speak of 'transfer' or 'transmission' in the context of *instantaneous* influences? "To speak of instantaneous travel from X to Y is a mixed or incoherent metaphor", van Fraassen [47] pointed out, ". . . for the entity in question is implied to be simultaneously at X and at Y—in which case there is no need for travel, for it is at its destination already." We agree with van Fraassen's assessment that—in the context of quantum nonlocality —the use of terms such as travel, transfer, propagation, or transmission, is logically incoherent. However, for the present study we have retained the use of such terms for historical reasons, and for reasons of scientific convention. That is, in relation to the notion of 'instantaneous influences', instances of "mixed or incoherent metaphor" abound in the literature on quantum foundations such as, to name only two examples, "propagation with infinite velocity" [8] or "nonlocal information transferral" [48]. Of course, these authors were fully aware of the mixed status of such expressions, but alternative terms are unavailable—even now—that have standard use. To remedy the inconsistency, van Fraassen [47] proposed that "... one should say instead that the entity has two (or more) coexisting parts, that it is spatially extended." That proposition is also in line with an intuitive characterization of Reichenbach's principle of 'common cause' [49]. However, important theoretical obstacles remain in place there also as discussed by Cavalcanti and Lal [50].

Following van Fraassen [47], the entity of 'nonlocal quantum information' should be described as information which exists—at space-like separated locations—in "two ... coexisting parts" or as being "spatially extended". Even so, would the use of a more accurate description of the nonlocal entity alter in any way the findings of the present study? Our conclusions would remain as valid as before, because our analysis is indifferent to whether one views nonlocal information as a function of instant transmission, Reichenbach's 'common causation', or van Fraassens's 'spatial extension'. Significantly, an effective non-signalling theorem would be violated under the following condition: not only the entity of 'information' would have to be "spatially extended", or be manifested instantly by "two ... coexisting parts", but—similarly—the entity of the 'message' as well. However, as was illustrated in Fig. 4b and c, there exists no such necessary link between the presence of information and the presence of a message. In short, 'nonlocal information' does not automatically equal 'nonlocal communication' (see legend to Fig. 4). Future work might construct and use more appropriate terminology, yet the present conclusions do not depend on this.

### 6.4 Conclusions

This communication-theoretic study has demonstrated that an effective non-signalling theorem allows for nonlocal quantum information transfer yet—at the same time—effectively denies superluminal signalling and communication. While this study has addressed key points, the exploration of novel possibilities has just begun. Everyone agrees that no *final* judgement can yet be delivered concerning the compatibility between, for example, *non-relativistic* de Broglie–Bohm theory, and the theory of special relativity. Nevertheless, while there might still possibly exist reasons for why nonlocal information transfer may represent a physically-unrealistic proposition for quantum mechanics, we have shown that the danger of superluminal signalling and communication is not one of them.

**Acknowledgments** Work by Jan Walleczek at Phenoscience Laboratories (Berlin) is partially funded by the Fetzer Franklin Fund of the John E. Fetzer Memorial Trust. Work by Gerhard Grössing at the Austrian Institute for Nonlinear Studies (Vienna) is also partially funded by the Fetzer Franklin Fund of the John E. Fetzer Memorial Trust. The authors wish to thank Siegfried Fussy, Johannes Mesa Pascasio, Herbert Schwabl and Nikolaus von Stillfried for their valuable contributions in developing these concepts.

Found Phys**Open Access** This article is distributed under the terms of the Creative Commons Attribution 4.0 International License (http://creativecommons.org/licenses/by/4.0/), which permits unrestricted use, distribution, and reproduction in any medium, provided you give appropriate credit to the original author(s) and the source, provide a link to the Creative Commons license, and indicate if changes were made.## References

1. Bell, J.S.: On the Einstein Podolsky Rosen paradox. Physics **1**, 195–200 (1964)
2. Bell, J.S.: Bertlmann's socks and the nature of reality, in Speakable and Unspeakable in Quantum Mechanics, pp. 139–158. Cambridge University Press, Cambridge (1987)
3. Norsen, T.: J.S. Bell's concept of local causality. Am. J. Phys. **79**, 1261 (2011). arXiv:0707.0401 [quant-ph]
4. Eberhard, P.H.: Bell's theorem and the different concepts of locality. Nuov. Cim. B **46**, 392–419 (1978)
5. Bell, J.S.: de Broglie-Bohm, delayed-choice, double-slit experiment, and density matrix. Int. J. Quantum Chem. **14**, 155–159 (1980)
6. Bell, J.S.: On the impossible pilot wave. Found. Phys. **12**, 989–999 (1982)
7. Bell, J.S.: Speakable and unspeakable in quantum mechanics, in Speakable and Unspeakable in Quantum Mechanics, pp. 169–172. Cambridge University Press, Cambridge (1987)
8. Bell, J.S.: La nouvelle cuisine, in Speakable and Unspeakable in Quantum Mechanics. Revised ed. Cambridge University Press, Cambridge (2004)
9. Bohm, D.: A suggested interpretation of the quantum theory in terms of "hidden" variables. I. Phys. Rev. **85**, 166–179 (1952)
10. Bohm, D.: A suggested interpretation of the quantum theory in terms of "hidden" variables. II. Phys. Rev. **85**, 180–193 (1952)
11. Shannon, C.E.: A mathematical theory of communication. Bell Syst. Tech. J. **27**(379–423), 623–656 (1948)
12. Shannon, C.E., Weaver, W.: The Mathematical Theory of Communication. The University of Illinois Press, Urbana (1949)
13. Scarani, V., Gisin, N.: Superluminal influences, hidden variables, and signaling. Phys. Lett. A **295**, 167–174 (2002)
14. Scarani, V., Bancal, J.-D., Suarez, A., Gisin, N.: Strong constraints on models that explain the violation of Bell inequalities with hidden superluminal influences. Found. Phys. **44**, 523–531 (2014)
15. Gallego, R., Masanes, L., De La Torre, G., Dhara, C., Aolita, L., Acín, A.: Full randomness from arbitrarily deterministic events. Nat. Commun. **4**, 2654 (2013). arXiv:1210.6514 [quant-ph]
16. Harrigan, N., Spekkens, R.W.: Einstein, incompleteness, and the epistemic view of quantum states. Found. Phys. **40**, 125–157 (2010)
17. Colbeck, R., Renner, R.: Is a system's wave function in one-to-one correspondence with its elements of reality? Phys. Rev. Lett. **108**, 150402 (2012)
18. Pusey, M. F., Barrett, J., Rudolph, T.: On the reality of the quantum state. Nat. Phys. **8** 475–478 (2012). arXiv:1111.3328
19. Ghirardi, G., Romano, R.: About possible extensions of quantum theory. Found. Phys. **43**, 881–894 (2013)
20. Ghirardi, G., Romano, R.: Ontological models predictively inequivalent to quantum theory. Phys. Rev. Lett. **110**, 170404 (2013)
21. Leifer, M.S.: Is the quantum state real? An extended review of $\psi$-ontology theorems. Quanta **3**, 67–155 (2014). arXiv:1409.1570
22. Wood, C.J., Spekkens, R.W.: The lesson of causal discovery algorithms for quantum correlations: causal explanations of bell-inequality violations require fine-tuning. New J. Phys. **17**, 033002 (2015). arXiv:1208.4119 [quant-ph]
23. Walleczek, J., Grössing, G.: The non-signalling theorem in generalizations of Bell's theorem. J. Phys. Conf. Ser. **504**, 012001 (2014). arXiv:1403.3588 [quant-ph]
24. Colbeck, R., Renner, R.: No extension of quantum theory can have improved predictive power. Nat. Commun. **2**, 411–415 (2011)
25. Colbeck, R., Renner, R.: Free randomness can be amplified. Nat. Phys. **8**, 450–454 (2012). arXiv:1105.3195 [quant-ph]
1820   Springer